\documentclass{JHEP} 

\usepackage{epsfig,multicol}

\def\ds{{\sffamily DarkSUSY}{\ }}

\newcommand{\code}[1]{{\tt #1}}

\newcommand\fverb{\setbox\pippobox=\hbox\bgroup\verb}
\newcommand\fverbdo{\egroup\medskip\noindent%
			\fbox{\unhbox\pippobox}\ }
\newcommand\fverbit{\egroup\item[\fbox{\unhbox\pippobox}]}
\newbox\pippobox
\newcommand{\beq}{\begin{equation}}
\newcommand{\eeq}{\end{equation}}

\def\neu1{\widetilde{\chi}^0_1}

\def\simlt{\stackrel{<}{{}_\sim}}

\newcommand{\wtld}[1]{\widetilde{#1}}
\def\neu1{\widetilde{\chi}^0_1}

\def\bu{\widetilde{B}_1}
\def\bd{\widetilde{B}_2}
\def\huz{\widetilde{H}_1^0}
\def\hdz{\widetilde{H}_2^0}

\title{ Dark Matter in split extended  supersymmetry}

\author{Alessio Provenza \\
	SISSA/ISAS, via Beirut 4, I-34013 Trieste, Italy, and\\
          INFN, Sezione di Trieste, I-34014 Trieste, Italy\\
	E-mail: \email{provenza@sissa.it}}

\author{Mariano Quiros \\
	Instituci\'o Catalana de Recerca i Estudis Avan\c{c}ats (ICREA) \\
        Theory Physics Group, IFAE/UAB,
        E-08193 Bellaterra, Barcelona, Spain \\
	E-mail: \email{quiros@ifae.es}}
	
\author{Piero Ullio \\
	SISSA/ISAS, via Beirut 4, I-34013 Trieste, Italy, and\\
          INFN, Sezione di Trieste, I-34014 Trieste, Italy \\
	E-mail: \email{ullio@sissa.it}}

\preprint{SISSA-46/2006/EP\\UAB-FT-607}

\abstract{We consider the split extended ($N=2$) supersymmetry
 scenario recently proposed by Antoniadis {\it et
 al.}~[hep-ph/0507192] as a realistic low energy framework arising
 from intersecting brane models. While all scalar superpartners and
 charged gauginos are naturally at a heavy scale, the model low energy
 spectrum contains a Higgsino-like chargino and a neutralino sector
 made out of two Higgsino and two Bino states. We show that the
 lightest neutralino is a viable dark matter candidate, finding
 regions in the parameter space where its thermal relic abundance
 matches the latest determination of the density of matter in the
 Universe by WMAP. We also discuss dark matter detection strategies
 within this model: we point out that current data on cosmic-ray
 antimatter already place significant constraints on the model, while
 direct detection is the most promising technique for the
 future. Analogies and differences with respect to the standard split
 SUSY scenario based on the Minimal Supersymmetryc Standard Model are illustrated.}


\begin{document} 

\section{Introduction}

In recent years cosmological observations have provided increasingly
convincing evidence that non-baryonic dark matter (DM) is the building
block of all structures in the
Universe~\cite{Spergel:2006hy}. Consequently when defining standard
model (SM) extension candidates of elementary particles, the
possibility to embed a DM candidate has become a compelling
guideline. In such respect the formulation of split
supersymmetry~\cite{Arkani-Hamed:2004fb,Giudice:2004tc} is no
exception.

Split supersymmetry (SUSY) labels a generic realization of a SUSY SM
extension with a "split" superpartner spectrum. On one side all
sfermions are assumed to be very heavy (at some intermediate scale
between, say, 100 TeV and the GUT scale): this feature explains why
SUSY contributions to flavor and CP violation are small at the cost of
invoking some mechanism, not related to SUSY, to stabilize the weak
scale~\cite{Arkani-Hamed:2004fb}, but still allowing for a successful
unification of gauge couplings. On the other hand (at least) some of
the fermionic superpartners need to be light: the cosmological
measurements of the matter density set an upper bound to the thermal
relic abundance of the lightest neutralino, the lightest
SUSY  particle (LPS)  in this framework, which in turn can be translated into
an upper bound on the LSP mass (about 340~TeV for a generic thermal relic~\cite{Griest:1989wd},  about 2.1~TeV for a pure Wino-like
neutralino, see e.g.~\cite{mpu}). The split SUSY framework is thus the
minimal SUSY setup which can accommodate a DM candidate: taking as
strong prior the condition that the LSP accounts for all DM in the
Universe sets tight constraints on the model.

A large class of scenarios predict a split SUSY
spectrum~\cite{Arkani-Hamed:2004yi,
Antoniadis:2004dt,Kors:2004hz,Babu:2005ui,Dutta:2005zz}. We will focus
here on the model arising in a string-inspired framework with
intersecting branes, recently introduced in Refs.~\cite{splitn2}. This
model has gauge and Higgs sectors defined in multiplets of extended
supersymmetry. Its low energy spectrum cannot be described as a subset
of the spectrum in the minimal SUSY SM extension (MSSM), which is the
standard lore for most phenomenological studies of low energy SUSY
models.  We will consider the LSP relic density calculation as a
guideline to examine the structure of the model, and discuss the
perspectives of testing neutralino DM in this scenario. Previous
studies of DM in split SUSY have assumed the MSSM as working
framework, see
e.g.~\cite{Pierce:2004mk,Arvanitaki:2004df,mpu,Senatore:2004zf,
Arkani-Hamed:2006mb} and so we will point out differences and
analogies with respect to the present case.

The structure of the paper is as follows. In Section~\ref{Sec:model}
we introduce the framework and discuss its low energy limit. In
Section~\ref{Sec:omega} we compute the LSP relic density.  We then
discuss in Section~\ref{sec:phenoguide} current constraints on the
model from direct and indirect DM searches, and prospects to test it
at future facilities. Section~\ref{Sec:concl} summarizes our results.

\section{The split extended SUSY framework}
\label{Sec:model}

We consider a low energy model arising from a high energy intersecting
brane model with split $N=2$ supersymmetry as discussed by Antoniadis
{\it et al.}~in Ref.~\cite{splitn2}.  At the electroweak scale the
active degrees of freedom are: the Standard Model ones with  the
usual SM Higgs sector replaced by a two Higgs doublet sector,  and an 
additional fermionic  sector made up by  two neutral and one charged Higgsino
states, as in a standard two Higgsino doublet structure, and two
(rather than one as in the MSSM) neutral states with Bino quantum
numbers, which are the  $N=2$ SUSY fermionic partners of the $U(1)_Y$ gauge field.         
The model does not contain light superpartners of the $SU(3)\times
SU(2)$ gauge fields (again in contrast with the standard MSSM setup),
as well as any light superpartner of the SM fermions (as in any split
SUSY framework). The four neutralino states are then obtained as mass
eigenstates from the superposition of two neutral Higgsinos and the
two Binos; the lightest neutralino is always the lightest SUSY
particle and, restricting ourselves to models with conserved R-parity,
stable. Analogously to more standard scenarios, since the LSP is
massive, stable, electric- and color-charge neutral, it is a natural
DM candidate.

As first step to study the phenomenology of the LSP as DM candidate, we
need to re-derive the spectrum and couplings in the present case.  We
recall that the general $N=2$ supersymmetric and gauge invariant
Lagrangian can be written in the $N=1$ superfield formalism
as~\footnote{For a review see for example~\cite{alvarez}.}
\begin{eqnarray}
\mathcal{L}&=&\mathcal{L}_{Kin.gaug}+\int
d^4\theta(H_1^{\dagger}e^{-2gV}H_1+H_2e^{2gV}H_2^{\dagger})\nonumber\\
&+&\int
d^2\theta\left(\mu H_2H_1+ g\sqrt{2}H_2\Phi H_1 \right)+h.c.
\label{N2lagr}
\end{eqnarray}
where $V=T^A V^A$ is the $N=1$ vector multiplet, $T^A$ being the
generators of the gauge group, and similarly $\Phi=T^A \Phi^A$ where
$\Phi$ is the chiral multiplet in the adjoint representation, the
$N=2$ partner of $V$. $H_1$ and $H_2$ are the two chiral multiplets
contained in the Higgs hypermultiplet; they transform as doublet and
anti-doublet, respectively, under $SU(2)$~\footnote{Notice the
different notation with respect to the MSSM where $H_1$ and $H_2$
stand for $SU(2)$ doublets.}.

From the previous equation it is straightforward to write the Higgs
potential which, due the presence of the last term in
Eq.~(\ref{N2lagr}), is different from the MSSM one. Taking into
account the soft-SUSY breaking terms it has the form:
\begin{eqnarray}
V&=&m_1^2|H_1|^2+m_2^2|H_2|^2+m_3^2(H_2H_1+h.c.)
          +{g^2\over8}\left(H_1^{\dag}\vec{\sigma}H_1
          - H_2\vec{\sigma}H_2^{\dag} \right)^2\nonumber\\
  &+&{g'^2\over8}\left(|H_2|^2-|H_1|^2\right)^2
          +{g'^2\over2}\left|H_2H_1\right|^2
          +{g^2\over2}\left|H_2\vec{\sigma}H_1\right|^2.\label{Hpot}
\end{eqnarray}
It is interesting to note that in this potential the so called D-flat
directions are absent and hence we do not need to put any extra
constraints on the quadratic terms coefficients of the potential to
avoid unbounded-from-below directions. This is entirely due to the
last term in Eq.~(\ref{Hpot}) arising from the $N=2$ structure of the
model.

The next step is to work out the scalar potential of the neutral field
in order to find a suitable configuration for the electroweak symmetry
breaking.  After the minimization we are left with two would-be
Goldstone bosons: the neutral one
$G^0=-\cos\beta\,Im[H^0_1]+\sin\beta\,Im[H^0_2]$ and the charged one
$G^+=-\cos\beta(H_1^-)^*+\sin\beta(H_2^+)$; a massive neutral CP-odd
particle $A^0=\sin\beta\, Im[H^0_1]+\cos\beta\, Im[H^0_2]$ with mass
$m_A$, two massive neutral CP-even Higgs bosons namely the SM like
$h=\cos\beta\, Re[H^0_1]+\sin\beta\, Re[H^0_2]$ with $m_{h}=m_Z$, and
$H=\sin\beta\, Re[H^0_1]-\cos\beta\, Re[H^0_2]$ with $m_{H}=m_A$, and
a charged massive particle $H^+=\sin\beta(H_1^-)^*+\cos\beta(H_2^+)$
with $m_{H^+}^2=m_A^2+2m_W^2$.  The Higgs spectrum has then the same
composition as the MSSM but with different tree-level masses and
mixings. On the other hand we do not expect any differences in
radiative corrections compared to the MSSM: the couplings of the Higgs
fields with matter fields are unchanged since the latter are
introduced in $N=1$ representations.  In particular in the limit
$m_A\rightarrow\infty$ we can apply the same expressions computed in
Ref.~\cite{Giudice:2004tc}.

From Eq.~(\ref{N2lagr}) we can also infer the masses for the SUSY
fermionic states.  In the $(\bu,\,\bd,\,\huz,\,\hdz)$ basis, the
neutralino mass matrix, including the soft-masses for Binos and
Higgsinos, takes the form:
\begin{equation}\mathcal{M}_{\wtld\chi^0}=
\left(
\begin{array}{cccc}
M & 0 & -m_Zs_Wc_{\beta} & m_Zs_Ws_{\beta}  
\\ 0 & M & m_Zs_Ws_{\beta}   & m_Zs_Wc_{\beta}\\
-m_Zs_Wc_{\beta}     & m_Zs_Ws_{\beta}  & 0 &-\mu \\ 
 m_Zs_Ws_{\beta}   &  m_Zs_Wc_{\beta}  & -\mu & 0
\end{array}
\right) 
\label{neumatrix}
\end{equation}
where $s_{\beta} \equiv \sin\beta$, $c_{\beta} \equiv \cos\beta$ and
$s_W \equiv \sin\theta_W$.  Eigenvector and eigenvalues of this matrix
can be analytically computed; the eigenvalues are~\cite{splitn2}:
\begin{equation}
m_{\chi_i}={1\over2}\left[(M+\epsilon_1\mu)-
\epsilon_2\sqrt{(M-\epsilon_1\mu)^2+4m_Z^2\sin^2\theta_W}\right]
\end{equation} 
where $\epsilon_i=\pm 1$.  The eigenvector for the lightest state, up
to a normalization factor, is:
$$
N_{1,i} \propto \left(-{2m_Z s_W(c_\beta+s_\beta)\over D(M,\mu)},
{2m_Z s_W(s_\beta-c_\beta)\over D(M,\mu)},-1,1\right). \label{eq:mixing}
$$
with $D(M,\mu) \equiv M-\mu+\sqrt{(M-\mu)^2+4m_Z^2\sin^2\theta_W}$.
Since the last two entries have the same modulus, the coupling of the
LSP with the $Z$ boson, which is proportional to
$|N_{1,3}|^2-|N_{1,4}|^2$, vanishes. The mass spectrum does not
depend on $\tan\beta$, which enters only in changing the relative
weight of the two Bino states for each neutralino [see the first two
entries in Eq.~(\ref{eq:mixing})]. The SUSY fermionic spectrum is
completed by one light chargino state, Higgsino-like, with mass
$m_{\chi^+} = \mu$.

The last preliminary step would be to derive the Feynman rules for
neutralinos, chargino and Higgses.\footnote{A numerical package for the implementation of the model in the \ds code
if available upon request from the authors. In the package manual
the list of relevant Feynman rules is also provided.}  In our phenomenological analysis
we will restrict ourselves to a three dimensional parameter space,
scanning over the parameters $\mu$ and $M$, for a few values of
$\tan\beta$.  Having verified \textit{a posteriori} that the roles of
the CP-odd, the charged and the heavy CP-even Higgs bosons are
marginal for the phenomenology we are interested in, we will focus on
the decoupling limit $m_A>>m_W$; in such a limit the light CP-even
Higgs is SM-like and its mass depends weakly on the mass scale for
SUSY scalars~\cite{Giudice:2004tc}: in our model, the latter is set by
the grand unification constraint to be around $10^{13}\, {\rm GeV}$,
and hence we expect the SM-like Higgs bosons mass to be about
$m_{h}\sim 160\,{\rm GeV}$.

\section{The lightest neutralino as dark matter candidate}
\label{Sec:omega}

We compute the LSP thermal relic density by interfacing the particle
physics framework we have introduced into the \ds\
package~\cite{Gondolo:2004sc}. Such package allows for high accuracy
solutions of the Boltzmann equations describing thermal freeze out.
In particular, in computing thermally-averaged LSP pair annihilation
cross-sections we include systematically all kinematically allowed
final states (we remind the relic density scales, approximately, with
its inverse); eventual co-annihilation effects from SUSY states nearly
degenerate in mass with the LSP are included as well. The density
evolution equation is then solved numerically.  The estimated
precision on the value of the relic density we derive is, for a given
set of input parameters setting masses, widths and couplings, of the
order of 1\% or better.  We will compare the computed LSP relic density
with the latest determination of the CDM component of the Universe by
the WMAP experiment~\cite{Spergel:2006hy}: $\Omega_{CDM} h^2 =
0.110\pm 0.007$.

\FIGURE[t]{
\centerline{
\epsfig{file=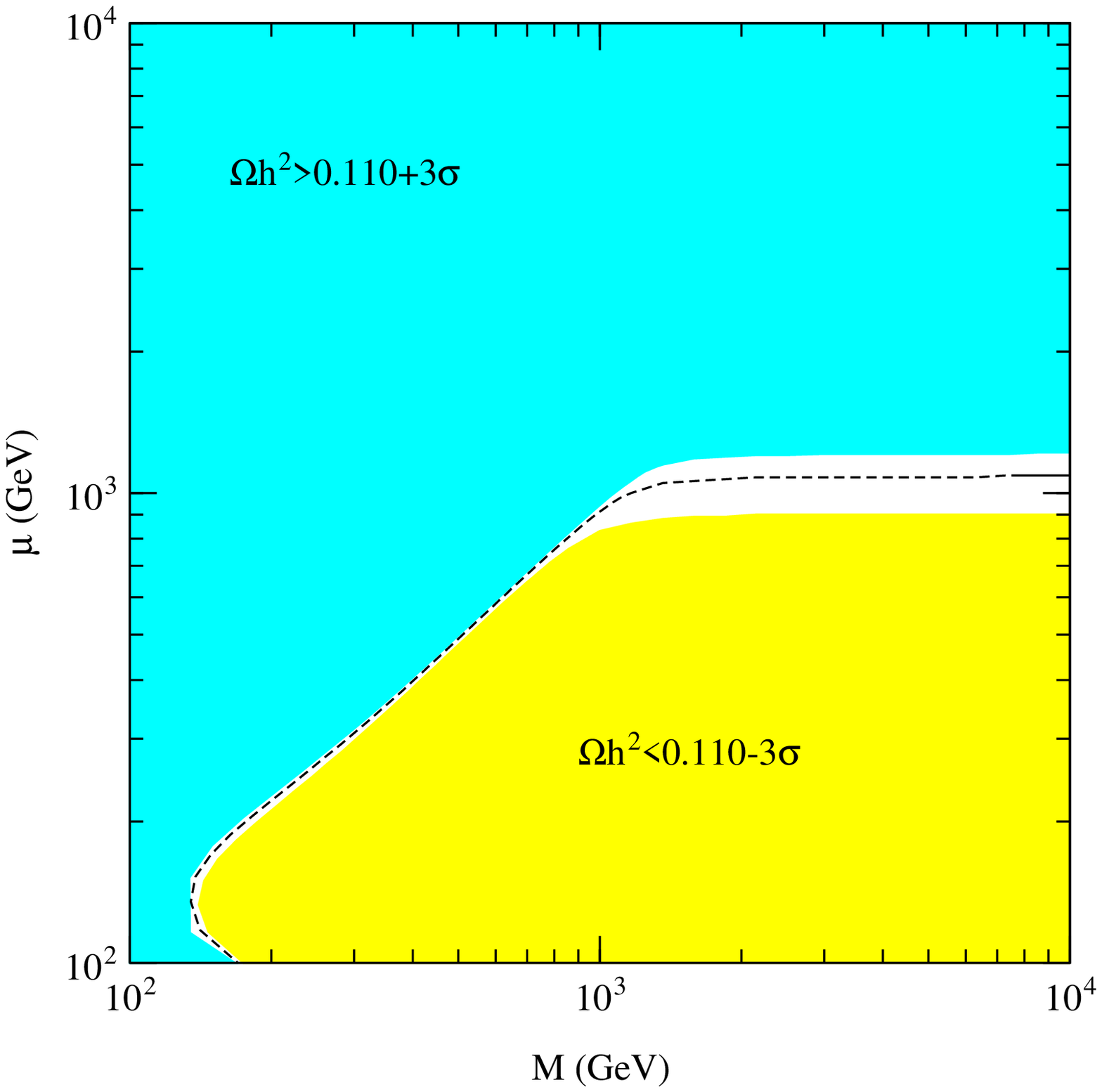,width=7.5cm}\quad
\epsfig{file=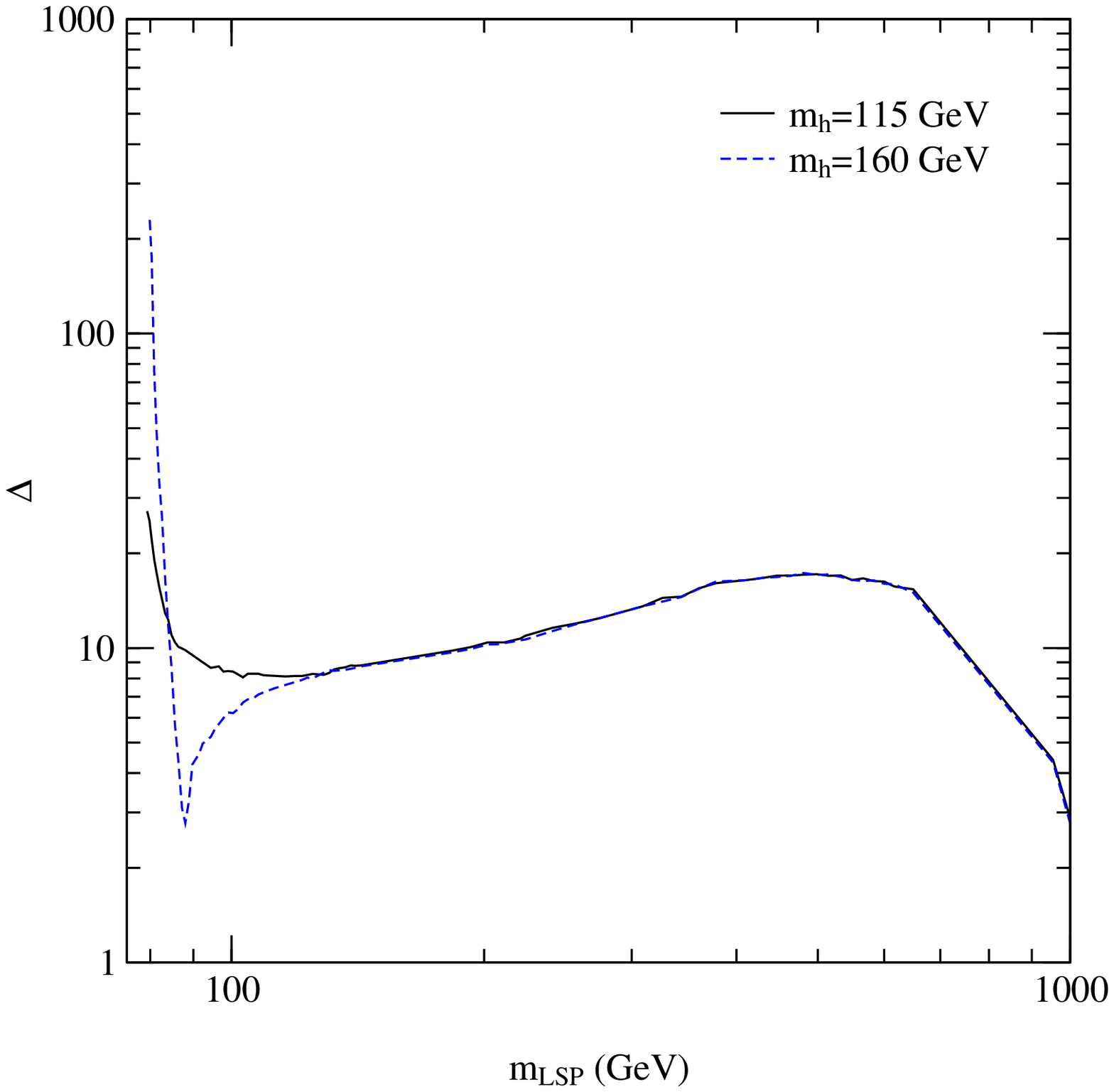,width=7.5cm}}
\caption{{\it Left Panel:} In white, the WMAP 3-$\sigma$ allowed
region in the plane $(M,\,\mu)$; the dashed line shows the isolevel
curve with $\Omega h^2=0.110$. $m_{h}$ is assumed to be equal
to 160~GeV.  \textit{Right Panel:} Fine-tuning parameter $\Delta$ versus
lightest neutralino mass for models with thermal relic abundance
$\Omega h^2=0.110$. The two sample values for $m_{h}$ are displayed,
featuring (or avoiding) an s-channel resonance in the LSP pair
annihilation rate.  }
 \label{fig:relic}}
%
In the left panel of Fig.~\ref{fig:relic} we plot the region of the
plane $(M,\,\mu)$ with relic density of the LSP matching the WMAP
preferred value for $\Omega h^2$ (0.110 along the dashed line and
within $3\sigma$ in the white region). There are two interesting
regimes: in the first one, at $M>>\mu$, the lightest neutralino is an
almost pure Higgsino and its thermal relic abundance is set by the
pair annihilation rate into gauge bosons, with coannihilations with
the next-to-lightest neutralino and the chargino playing an important
role. In this regime the LSP mass is essentially equal to $\mu$ and
imposing $\Omega h^2=0.110$, one finds $\mu \simeq 1.1~{\rm TeV}$.
The second interesting region starts at $M \sim \mu \sim 1~ {\rm TeV}$
and extends down to small $M$ and $\mu$, along the diagonal in the
plot. For these models the LSP has a large gaugino-Higgsino mixing,
while the mass splitting with the next-to-lightest neutralino and the
chargino gets larger. In this regime the thermal relic abundance is set by the pair annihilation rate into gauge bosons and coannihilation with chargino which 
get less and less relevant. For a given LSP mass, a tuning   on the LSP Higgsino fraction 
(and consequently on the LSP pair annihilation rate into gauge bosons) is  then needed to recover 
the correct relic abundance.    The plot is
obtained for $\tan\beta = 30$ and $m_{h}\sim 160\,{\rm GeV}$.
Changing $\tan\beta$ does not affect out results, since it does not
change the Higgsino fraction.  The SM-like Higgs boson mediates the
s-channel annihilation of LSP's into a fermion-antifermion pair; this
process is subdominant except when close to the s-cahnnel resonance,
i.e.~for $m_{LSP} \sim m_{h}/2$. In the plot this takes place at a LSP
mass of $\sim$~80~GeV (the turnaround at the lower-left corner in the
plot); if we shift $m_{h}$ to a lower value, say the current limit on
the SM-like Higgs mass $\sim$~115~GeV~\cite{pdg}, the cosmologically
preferred region is essentially unchanged, since 80~GeV is anyhow the
threshold for $W$-boson final states, while at the resonant mass a
very tiny slice of the parameter space becomes allowed (the $h$-Higgs
boson has a very small width, about 0.3~GeV).  In the regime at large
Bino-Higgsino mixing, the $3\sigma$ WMAP-allowed region is very
narrow; in the right panel of Fig.~\ref{fig:relic} we plot, versus LSP
mass and for models with relic density $\Omega h^2=0.110$, the
fine-tuning parameter $\Delta$, defined as:
\begin{equation}
\Delta = \sqrt{\sum_{i} \left| d\log(\Omega h^2)\over d\log x_i\right|^2}
\end{equation}
where the sum is over the two $x_i$ parameters $\mu$ and $M$. Moderate
values of $\Delta$ (say $\simlt 10$) can be obtained. For pure
Higgsinos the fine tuning is very small, while it gets large for
models that have, at the same time, large mixing and relic density
dominated by coannihilation effects (largest sensitivity on the tuning
of the Higgsino fraction). The peak in $\Delta$ at the LSP low mass end
is due either to the $W$-boson threshold effect or, specially, to
the $h$ resonance.

In the discussion on DM detection rates we will make predictions also
for models that are outside the $3\sigma$ WMAP preferred region, still
under the assumption that they account for the dark matter in the
Universe. The reason for such a choice is that the results we have
shown are not totally general: a few assumptions on the particle
physics and cosmological model are involved and, if they are relaxed,
the value of the relic density may either increase or decrease. E.g.,
there could be extra non-thermal sources on top of the thermal
component; the Universe expansion rate at freeze out could be faster
than the value extrapolated according to the SM particle content, or
there could be injections of entropy at late times diluting the
thermal relic density component. Effects of this kind have been
discussed, e.g.~in~\cite{otheromega}.

\section{Detection rates}
\label{sec:phenoguide}

A very strong experimental effort is currently devoted to searches for
dark matter in the form of weakly interactive massive particles
(WIMPs), see, e.g.~Ref.~\cite{Bergstrom:1998hd} for a review. We
discuss here techniques which are relevant for the LSP in the present
framework. All predictions are obtained with an appropriate interface
of the model to the \ds package.

\subsection{Direct detection}
\FIGURE[t]{
\centerline{
\epsfig{file=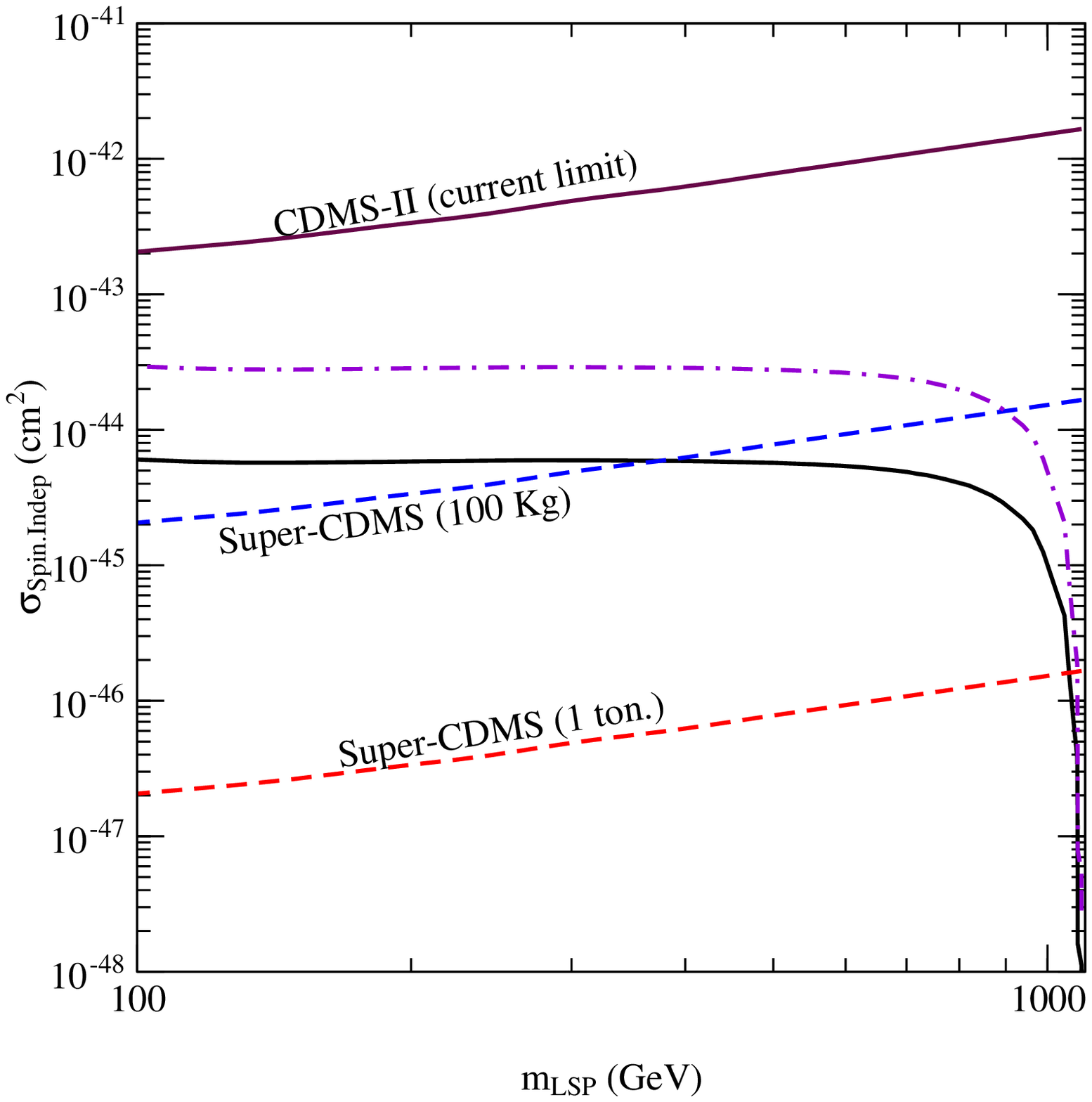,width=7.5cm} 
\caption{Lightest neutralino spin independent cross section per
nucleon versus lightest neutralino mass, for models with $\Omega h^2 =
0.110$. The solid line corresponds to the case of $m_{h} =
160\,\rm{GeV}$, while the dash-dotted line to models with $m_{h} =
115\,\rm{GeV}$. Also the current limit from the CDMS
experiment~\protect{\cite{cdms}} and future projected sensitivities
are shown.}\label{fig:ddonline}}}

The goal of WIMP direct searches is to measure the energy deposited
through elastic scatterings with nuclei by DM WIMP's passing through
the target material of a detector~\cite{dirdet}. Several complementary
approaches have been developed to optimize signal versus
backgrounds. In general for neutralinos, as for any Majorana fermion,
two terms can contribute to the scattering cross section: the
axial-vector spin-dependent (SD) coupling, and the scalar
spin-independent (SI) term, which is coherent and tends to dominate
for materials made out of heavy nuclei.  In particular in a split SUSY
scenario the LSP-nucleon SD and SI couplings can only be mediated by
the $Z$ boson and a CP-even Higgs in a t-channel, respectively, since
all squark contributions are suppressed by the fact that squarks are
very heavy. As already pointed out in Section~\ref{Sec:model}, at
tree-level, the vertex $Z$-LSP-LSP vanishes, hence the SD scattering
cross-section is always extremely small. In Fig.~\ref{fig:ddonline} we
present predictions for the SI neutralino-proton scattering
cross-section, versus lightest neutralino mass and for the model with
relic density $\Omega h^2 = 0.110$ (standard values for nucleonic
matrix elements are assumed, see~\cite{Gondolo:2004sc}). The results
are proportional to $1/m_{h}^4$ and they are sensitive to the
Bino-Higgsino mixing since the $h$-LSP-LSP vertex is proportional to it.
The conspiracy of the previous effects takes into account the sudden fall
of the SI cross section in Fig.~\ref{fig:ddonline}, having in mind that higher values
of LSP mass imply a small mixing between Binos and Higgsinos as shown in Fig.~\ref{fig:relic}. 
Also shown in Fig.~\ref{fig:ddonline} are the current best limit on
the WIMP-nucleon SI cross-section from the CDMS
collaboration~\cite{cdms} and projected sensitivities of upcoming
larger-mass detectors, such as the SuperCDMS~\cite{Akerib:2006rr}
(expectations for other projects with comparable masses are
analogous). No model in our framework is excluded by current
constraints; note however that a very large portion of the parameter
space will be tested at future facilities. This is even more evident
from the left panel of Fig.~\ref{fig:dd}, where future projected
sensitivities are plotted in the $\mu-M$ plane: the whole regime with
mass parameters below about 1~TeV will be probed (conservatively,
$m_{h}$ was set equal to 160~GeV).
%
%
\FIGURE[t]{
\centerline{
\epsfig{file=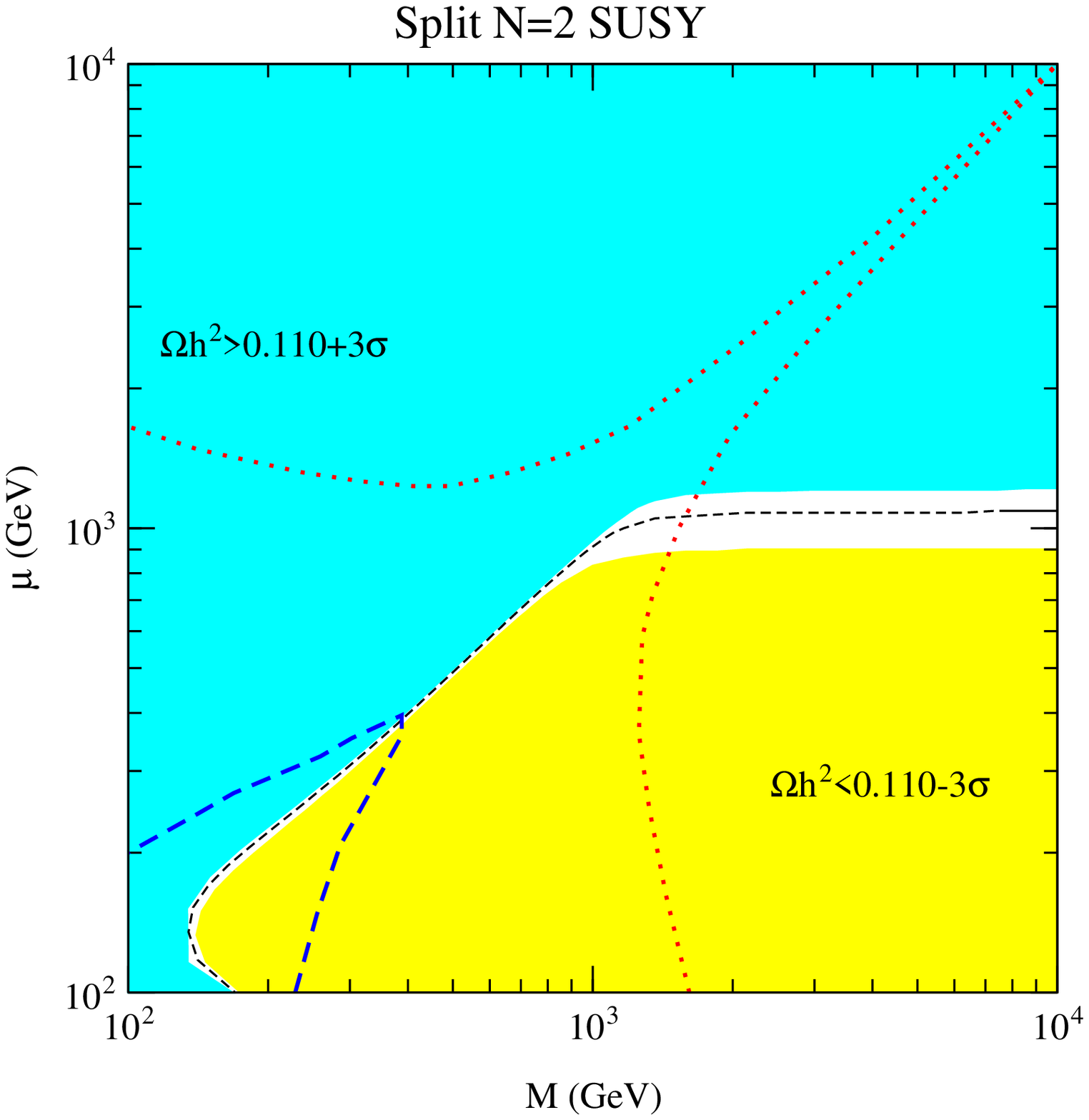,width=7.5cm}  \quad
\epsfig{file=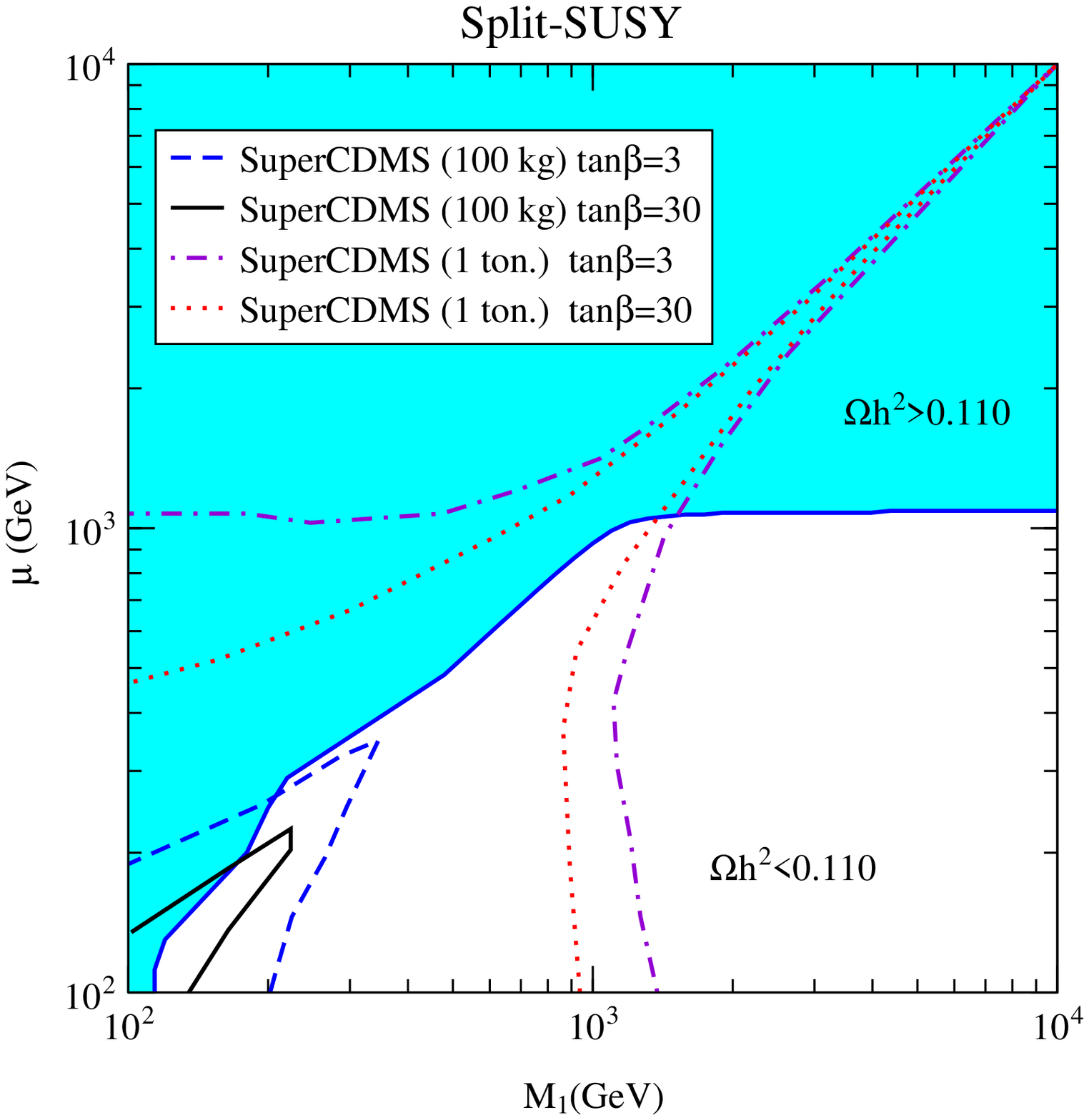,width=7.5cm}}
\caption{\textit{Left Panel:} Projected future sensitivities of the
SuperCDMS experiment in the $(M,\,\mu)$ plane for our split $N=2$ SUSY
model; the dotted line corresponds to the 1~ton configuration, the
dashed line to the 100 kg setup. Results are not sensitive to
$\tan\beta$.  \textit{Right Panel:} Sensitivity curves in the
$(M_1,\,\mu)$ plane for a split SUSY model within the standard MSSM
setup; two sample values of $\tan\beta$ are considered.  In both
frameworks $m_{h} = 160\,\rm{GeV}$ is assumed.}
\label{fig:dd}}

In the right panel of Fig.~\ref{fig:dd} we sketch the analogous
picture in the case of a split spectrum within the standard MSSM; here
$M_1$ is the mass parameter for the (single) bino, while it has been
assumed that the wino mass parameter $M_2$ is much heavier than $M_1$
and $\mu$ (see~\cite{mpu} for details).  While trends are similar,
there are a few significant differences: the departure from the
diagonal of the $\Omega h^2 = 0.110$ isolevel curve is due to the
$t\bar{t}$ final state which becomes kinematically allowed at $M_1
\sim \mu \sim 200$~GeV and it is mediated by a $Z$ boson in the
s-channel (the vertex $Z$-LSP-LSP is not zero in this case). Moreover
the SI scattering cross section depends on $\tan\beta$, since this
intervene in determining the LSP Bino-Higgsino mixing: from moderate to
large values of $\tan\beta$ the region in the parameter space
accessible to future searches shrinks, and it is in all cases less
extended than the corresponding region for our split extended SUSY
framework, which does not depend on $\tan\beta$ because of the
two-Bino construction.

\subsection{Indirect searches with neutrino telescopes}

Since WIMP's have a (small) coupling to ordinary matter, they can get
trapped in the potential well of massive celestial bodies, sink in the
center of the system and build up a dense WIMP population which in
turn can give rise to a neutrino flux. This effect has been studied in
detail for the Sun and the Earth~\cite{Bergstrom:1998hd} and, roughly
speaking, the trend one sees is that detectable fluxes are obtained
only for models for which capture and annihilation rates reach
equilibrium (or are close to it). 

In particular, the capture rate in the Sun is efficient if the SD
scattering cross section is sizable.
As already mentioned in our
framework the SD coupling is negligible and we find that the SI one
does not compensate for it. We find predictions for neutrino fluxes
which are always well below the expected sensitivity of future km$^3$
size telescopes, such as IceCube~\cite{Achterberg:2005fs}. This is in
contrast to what one finds for split models within the
MSSM~\cite{mpu}. In the latter case the SD coupling is not suppressed
and a detectable neutrino signal is expected for models with large
gaugino-Higgsino mixing and mass up to about 500~GeV; this could be in
principle exploited to discriminate between our framework and the
standard scenario.  In case of the Earth, the SI coupling is the
relevant effect; however we still find that the predicted neutrino
fluxes are too small.

\subsection{Indirect searches through cosmic- and gamma-rays}
\FIGURE[t]{
\centerline{
\epsfig{file=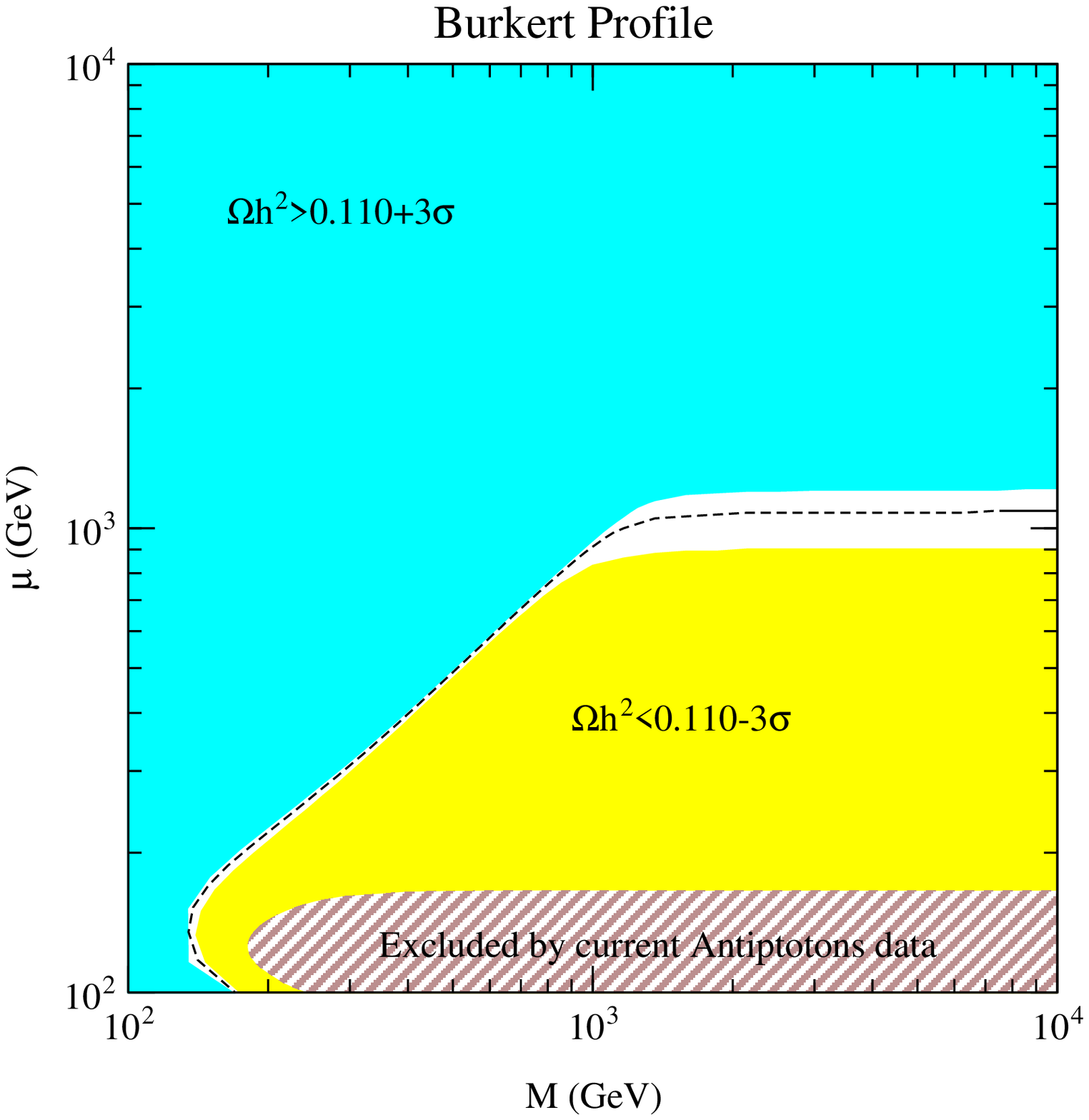,width=7.5cm}  \quad
\epsfig{file=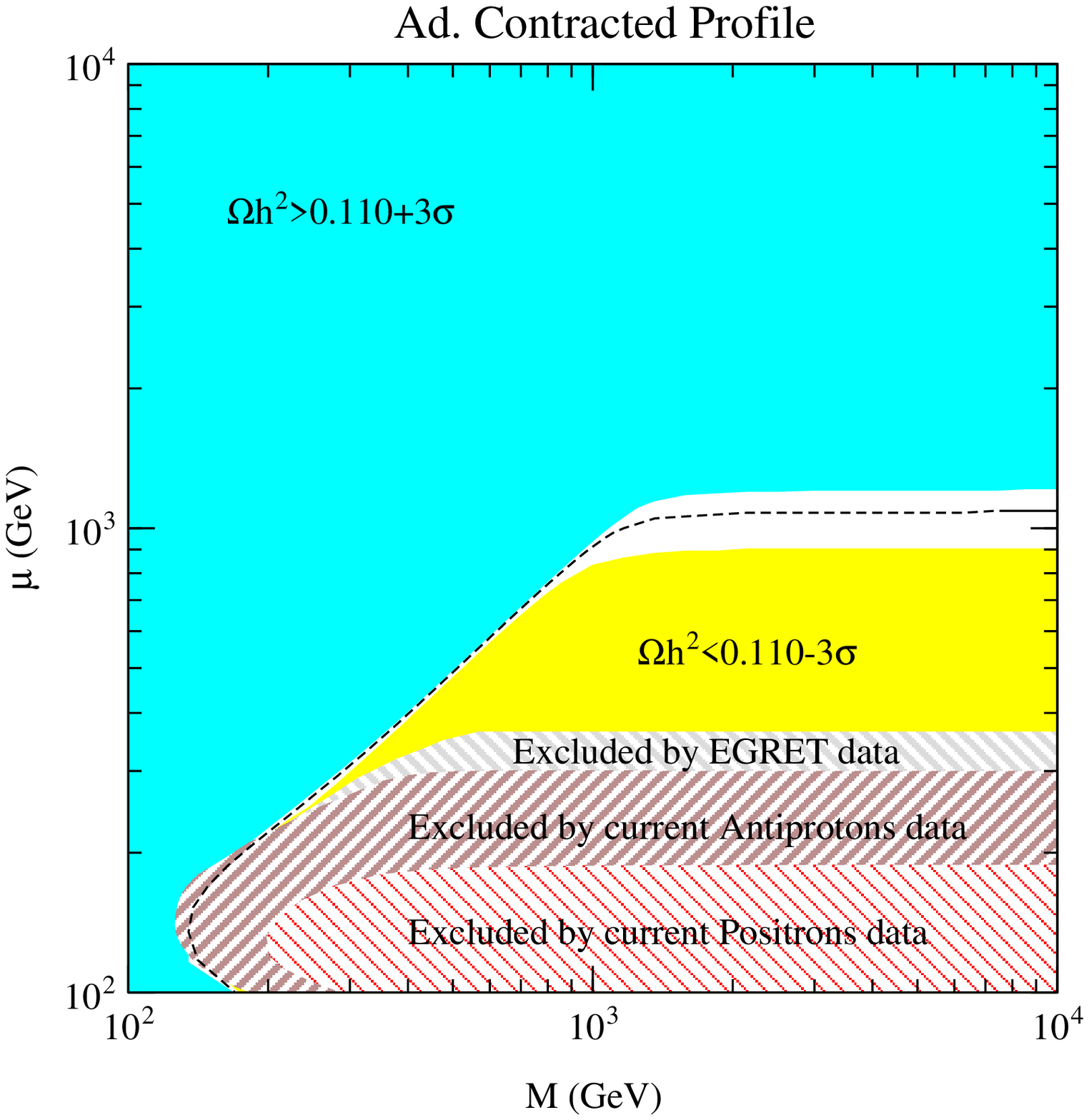,width=7.5cm}}
\caption{Limits on the model from the available measurements of the
local antimatter flux and from the gamma-ray flux measured by EGRET
towards the center of the Galaxy, in case of the Burkert halo profile
(left panel) and the adiabatically contracted profile (right
panel).}\label{fig:limits}}

If WIMP's are indeed the building blocks of all structures in the
Universe, they would populate the Milky Way halo as well, and their
pair annihilation could give rise to detectable yields.  The focus is
on species with small or well-determined background from standard
sources, such as antimatter components, i.e.~antiprotons, positrons
and antideuterons, and gamma-rays.  In order to make predictions for
the induced fluxes we need to simulate for the fragmentation and/or
decay processes for each final state of LSP annihilations to estimate
the energy distribution of these yields. This is done in the \ds
package by linking to simulations performed with the
\code{Pythia}~\cite{pythia} Monte Carlo code, except for the $\bar{D}$
source for which the prescription suggested in Ref.~\cite{dbar} is
implemented.

Since source functions scale with the square of the LSP density locally
in space, the induced fluxes will be sensitive to the distribution of
LSP's in the halo. We will refer to two opposite configurations: i) A
Burkert profile~\cite{burkert}, which has a large core and hence a
constant density in its inner region, essentially from our position in
the Galaxy inward; and ii) A profile matching the results from N-body
simulation of cold DM~\cite{n03}, i.e.~a profile with a sharp
enhancement in the density towards the center of the system, as
reshaped from the infall of the baryonic components of the Milky Way
in the adiabatic contraction limit~\cite{blumental}~\footnote{In this
second case the profile has essentially a $1/r^{1.5}$ singularity,
with cut-off in its innest 1~pc~\cite{ulliobh,milo}; more details on
the two profiles are given in Refs.~\cite{Edsjo:2004pf,mpu}, where the
same sample models are considered.}.  The spread in the predictions we
will show for the two configurations will give a feeling for the
uncertainty in the predictions related to the halo model. Even in the
case of the second model, we are not considering the most favorable
scenario; including, e.g.~effects due the presence of substructures
would give a further enhancement in the predictions.

To make predictions on antimatter fluxes one further step has to be
considered, in particular the simulation of the propagation of charged
particles in galactic magnetic fields. We use the two dimensional
diffusion model developed in Refs.~\cite{pbarpaper,epluspaper}, as
included in the \ds package, with a set of propagation parameters
which was shown to reproduce fairly well the ratios of primary to
secondary cosmic ray nuclei~\cite{strmosk} with the
\code{Galprop}~\cite{galprop} propagation code in the
diffusion/convection limit.  Solar modulation is instead sketched with
the analytical force-field approximation~\cite{GleesonAxford}, with a
modulation parameter as appropriate at each phase in the solar cycle
activity.

In Fig.~\ref{fig:limits} we show current limits on the split $N=2$
SUSY model from the available measurements of the local antiproton and
positron cosmic-ray fluxes~\cite{pbar,eplus}, for the two halo models
we are considering.  The limits are derived at $3\sigma$ with a
$\chi^2$ discrimination procedure, taking a standard prediction for
the background and neglecting uncertainties on it. For the Burkert
profile we do not have any constraint from positrons and gamma-rays
flux measurements while antiproton data do not exclude any model on
the $\Omega h^2 = 0.110$ isolevel curve. For the adiabatically
contracted case low mass models with thermal relic abundance in
agreement with WMAP data tend to give a too large antiproton flux
while also positron flux measurement rules out a corner in the plane
$(M,\mu)$. Furthermore, in this second case a part of the parameter
space is also excluded because it gives a gamma-ray flux which is
larger than the flux measured by the EGRET gamma-ray telescope towards
the Galactic Center~\cite{MH}.  More recent data at higher energy from
the HESS~\cite{hess} and MAGIC~\cite{magic} telescopes give less tight
constraints.

\FIGURE[t]{
\centerline{
\epsfig{file=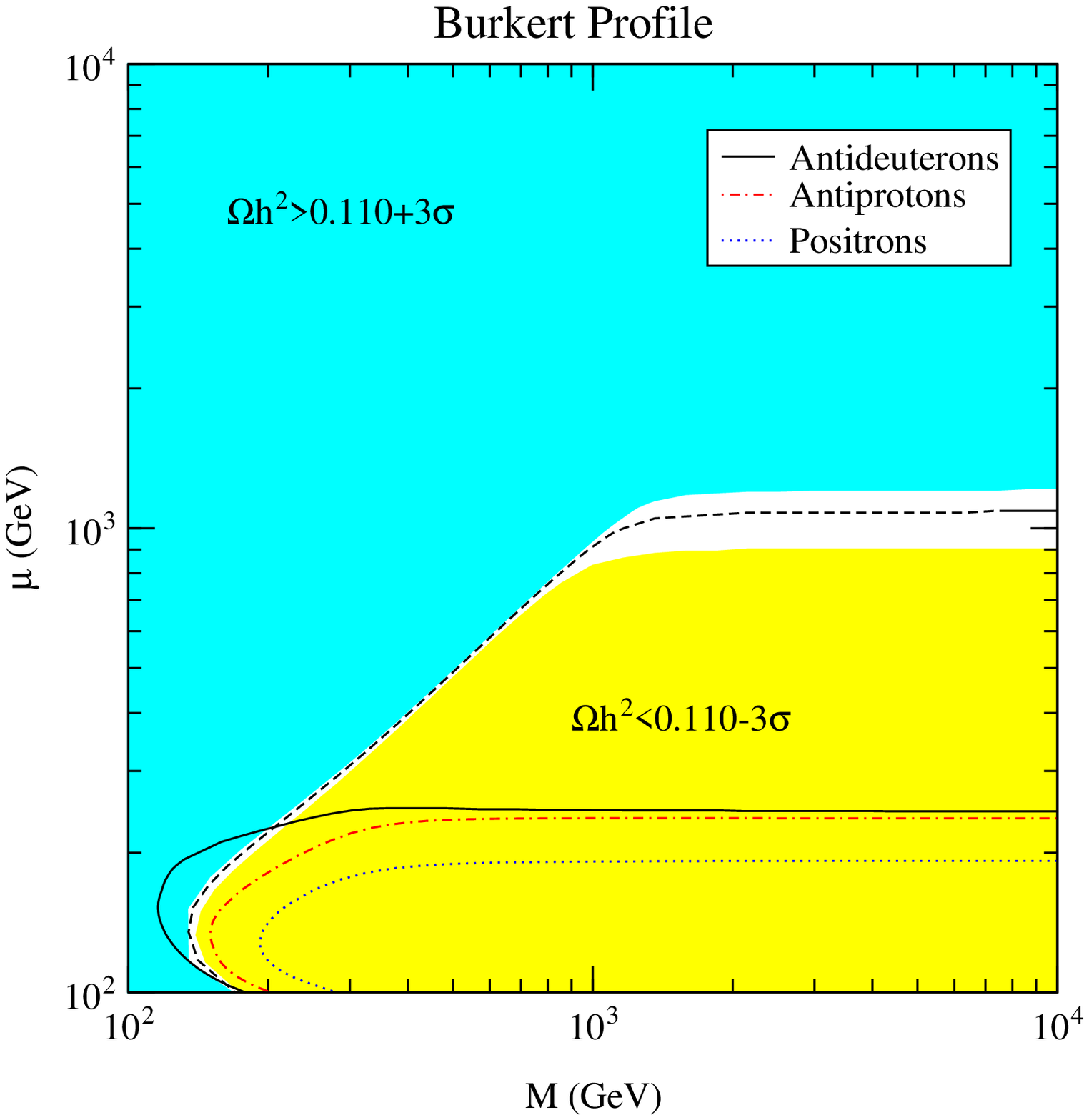,width=7.5cm} \quad
\epsfig{file=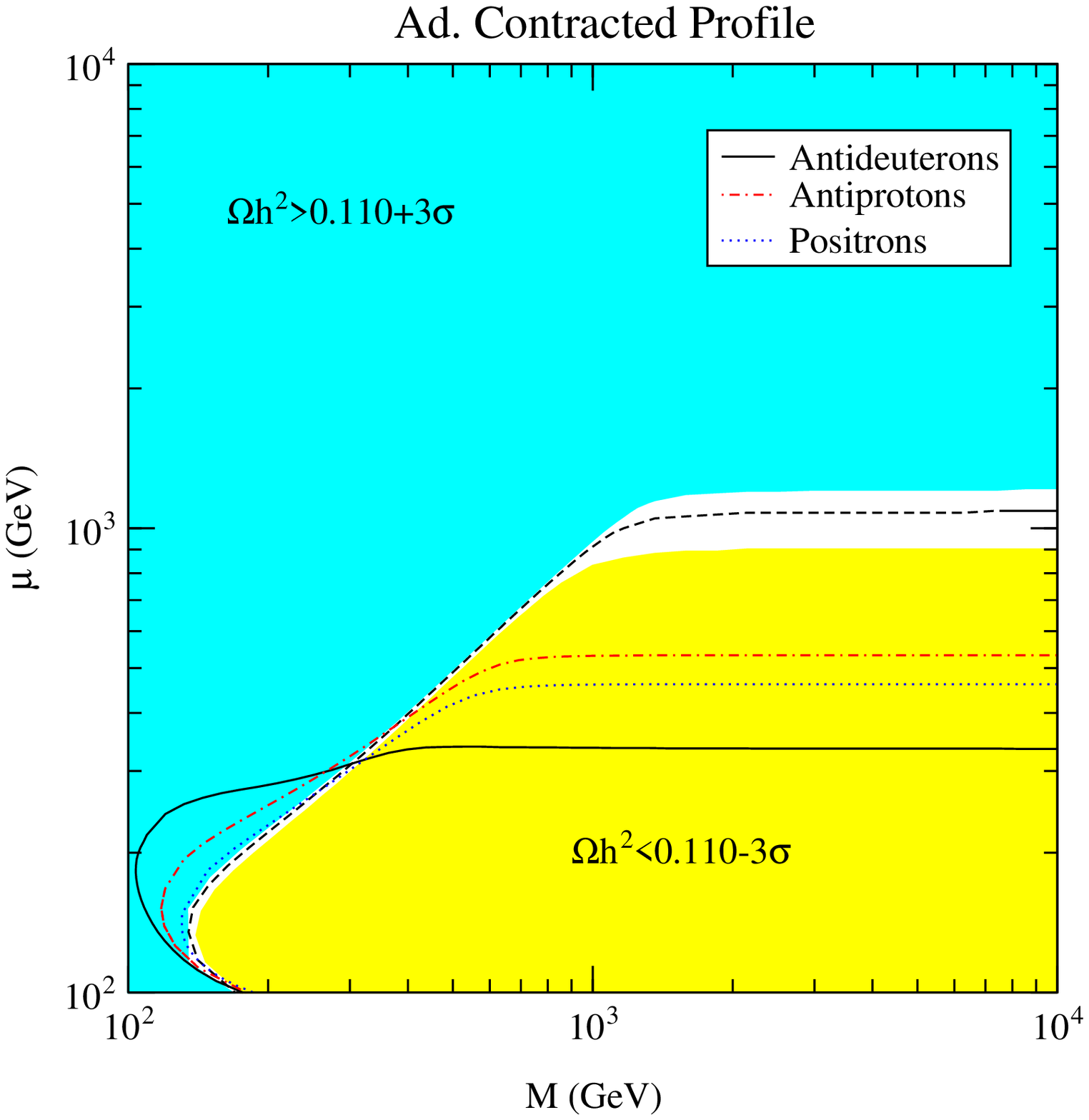,width=7.5cm}}
\caption{Portions of the parameter space which will be tested by the
Pamela and GAPS measurements of the antiproton, positron and
antideuteron fluxes in case of the Burkert halo profile (left panel)
and the adiabatically contracted profile (right panel).}
\label{fig:indd}}
Very recently the PAMELA satellite experiment~\cite{pamela} has been
launched. In the next few years very high quality data on the
antiproton and positron fluxes will be available, allowing for a
better discrimination of a component from WIMP annihilations.  In
Fig.~\ref{fig:indd} we show the regions in the parameter space which
will be tested by Pamela in three years of data taking (the
extrapolation on these sensitivity curves is done following the
approach described in Ref.~\cite{stefanopiero}).  We also show the
prospects for indirect detection with antideuteron searches obtained
by comparing the predicted flux with the estimated sensitivity of the
planned gaseous antiparticle spectrometer (GAPS)~\cite{Mori}, about
$2.6\times10^{-9}\textrm{m}^{-2}\textrm{sr}^{-1}\textrm{GeV}^{-1}
\textrm{s}^{-1}$ in the 0.1-0.4~GeV energy range~\footnote{We have
considered a configuration for the instrument placed on a satellite on
earth orbit.  The idea of an instrument on a deep space probe has been
considered as well, a case which would be more favorable for DM
detection.}. All indirect signals scale with the pair annihilation
cross section, which is large for pure Higgsinos and get very small
for pure Binos; it is inversely proportional to the square of the LSP
mass. The dominant annihilation channels are gauge boson final states,
which are copious sources of antiprotons and positrons, and which tend
to give LSP signals with very distinctive features, see e.g.~the
"W-boson bump" discussed in~\cite{epluspaper} for the positron
flux. The shift in the sensitivity curves going from the case with LSP
distribution described by the Burkert profile to the adiabatically
contracted case is very large and differs in the three detection
channels; this is due to the fact that spatial propagation is rather
efficient for antiprotons, while antideutrons are more fragile and
positrons lose energy on a shorter timescale. The relative weight of
the three states is also changing with the value of LSP mass, with the
antideuteron yield getting suppressed for heavier neutralinos.

Finally, comparing the perspectives for halo signals with those for
direct detection in Fig.~\ref{fig:dd}, one can see that there is
complementarity between the region of the parameter space probed by
indirect detection and the one accessible to direct searches; in this
respect, the result of this analysis is analogous to what was found
for split SUSY models within the MSSM framework, see~\cite{mpu}.

\section{Conclusion}
\label{Sec:concl}

We have discussed the dark matter phenomenology for a split SUSY model
arising from a high energy intersecting brane model with $N=2$
supersymmetry. Its active states at low energy differ from those in
the standard split SUSY scenario based on the MSSM.  Analogies and
differences compared to the standard case have emerged. In both
frameworks the lightest neutralino is a dark matter candidate for: i)
Intermediate mass neutralinos with a sizable Bino-Higgsino mixing, or
ii) A pure Higgsino state at the TeV scale. The fine-tuning parameter
for such configurations is in general rather small. In the split $N=2$
SUSY model direct detection is very promising, covering a
significantly larger portion of the parameter space than in the
standard case; in particular all results in the present analysis are
essentially independent of $\tan\beta$, and the spin independent
scattering cross section does not get suppressed in the large
$\tan\beta$ limit as in the MSSM. Concerning indirect detection
searches with neutrino telescopes are not relevant in this specific
extended SUSY scenario, while antimatter measurements have already
excluded a relevant portion of the parameter space. In the next few
years the Pamela detector will measure the antiproton and positron
fluxes with higher accuracy, allowing for further tests of the model
in a regime which is complementary to direct detection searches.

\section*{Acknowledgments}

The work of A.P.\ and P.U.\ was supported by the Italian INFN under
the project ``Fisica Astroparticellare'' and the MIUR PRIN ``Fisica
Astroparticellare''.
The work of M.Q. was partly supported by CICYT,
Spain, under contracts FPA 2004-02012 and FPA 2005-02211,
and partly by the European Union through the Marie
Curie Research and Training Networks "UniverseNet"
(MRTN-CT-2006-035863) "Quest for Unification" (MRTN-CT-2004-503369).

\end{document}